\documentclass[prd,showpcs,amsmath,amssymb,nofootinbib,longbibliography,twocolumn,showpacs,notitlepage,superscriptaddress]{revtex4-1}

\usepackage[dvips]{color}
\usepackage[utf8]{inputenc}

\usepackage{latexsym}
\usepackage{epsfig}
\usepackage{graphicx,subfigure}
\usepackage[normalem]{ulem}
\usepackage{color}
\usepackage{xcolor}
\usepackage{multirow}
\usepackage{hyperref}
\usepackage{numprint}
\usepackage{eucal}
\usepackage{amsmath}
\usepackage{amssymb}

\usepackage{xcolor}

\newcommand{\Lg}{\mathcal{L}}

\newcommand{\h}{\mathcal{H}}

\newcommand{\ssout}[1]{}

\def\ga{\mathrel{\raise.3ex\hbox{$>$\kern-.75em\lower1ex\hbox{$\sim$}}}}
\def\la{\mathrel{\raise.3ex\hbox{$<$\kern-.75em\lower1ex\hbox{$\sim$}}}}

\def\be{\begin{equation}}
\def\ee{\end{equation}}
\def\bea{\begin{eqnarray}}
\def\eea{\end{eqnarray}}

\def\T{\rm T}
\def\E{\rm E}
\def\J{\rm J}
\def\L{\rm L}

\begin{document}

\preprint{CERN-TH-2023-017}
\begin{flushright}
KCL-PH-TH-2023-08 \\
CERN-TH-2023-017
\end{flushright}

\title{The mirage of luminal modified  gravitational-wave  propagation}
\author{Antonio Enea Romano}
\affiliation{Theoretical Physics Department, CERN, CH-1211 Geneva 23, Switzerland}
\affiliation{ICRANet, Piazza della Repubblica 10, I--65122 Pescara, Italy}
\affiliation{Instituto de Fisica,Universidad de Antioquia,A.A.1226, Medellin, Colombia}

\author{Mairi Sakellariadou}
\affiliation{Theoretical Particle Physics and Cosmology Group,\\ Physics Department,
King’s College London, University of London, Strand, London WC2R 2LS, UK}

\date{\today}

\begin{abstract}
Using conformal invariance of gravitational waves, we show that for a luminal modified gravity theory, the gravitational-wave propagation and luminosity distance are the same as in general relativity.
The relation between the gravitational-wave and electromagnetic-wave luminosity distance gets however modified for electromagnetism minimally coupled to the Jordan frame metric. 
Using effective field theory we show that the modified relation obtained for luminal theories is also valid for non-luminal theories with Jordan frame matter-gravity coupling.
We generalise our analysis to a time-dependent speed of gravitational waves with matter minimally coupled to either the Jordan or Einstein frame metrics.
\end{abstract}

\pacs{Valid PACS appear here}

\maketitle

\textbf{Introduction---}
With the detection of gravitational waves (GWs) \cite{LIGOScientific:2016aoc} by the Laser Interferometer Gravitational Wave Observatory (LIGO) and Virgo, we entered the era of gravitational multi-messenger astronomy. These observations are in good agreement with general relativity (GR) predictions,   constraining modified gravity theories (MGTs) and dark energy (DE) models. Bright sirens, namely GW events with an electromagnetic counterpart \cite{LIGOScientific:2017vwq}, are used to test MGTs \cite{Baker:2017hug,Creminelli:2017sry,Sakstein:2017xjx,Ezquiaga:2017ekz,Wang:2017rpx} which predict a difference between the speed $c$ of electromagnetic waves (EMWs) and the speed $c_{\T}$ of gravitational waves.
The event GW170817 led to tight constraints on the difference between $c_{\T}$ and $c$, motivating  investigation of MGTs with $c_{\T}=c$ \cite{Belgacem:2018lbp}.

We show that for any MGT with $c_{\T}=c$, the GW propagation is the same as in GR, threrefore any apparent modification of the friction term of the GW propagation equation can be removed by switching from the Jordan frame, customary used to study MGTs, to the Einstein frame.  As a result, the choice $c_{\T}=c$ implies that the GW luminosity distance is the same as in GR, and if  matter is minimally coupled to the Einstein frame metric, the GWs and EMWs luminosity distances are the same.  
If however matter is non-minimally coupled to the metric in the Einstein frame, the GW and EMW luminosity distances can be different. 
In other words, GW do not feel the effective Planck mass associated to the conformal transformation between Jordan and Einstein frames, while photons, and in general matter fields, feel it.
Using  effective field theory we derive the general model-independent relation between GW and EMW luminosity distances for MGTs with a time-dependent $c_{\T}$, for matter fields coupled to either the Einstein or the Jordan frame metrics, generalising results obtained %previously 
assuming $c_{\T}=c$.

%%%%%%%%%%
\textbf{Apparent modification  of GW propagation equation in GR---}
 The Lagrangian of tensor modes in GR in an expanding Universe is
\bea
\Lg_{\rm GR}&=&a^2_{\E}\Big[ h'^2- (\nabla h)^2\Big]\ ,\label{LGR}
\eea
which gives the equation of motion 
\bea
h''+2 \h_{\E} h'-\nabla^2 h=0\,,
\label{GWGR}
\eea
where $\h_{\E}=a_{\E}'/a_{\E}$.

The use of conformal time makes it 
transparent to understand the effects of a time-dependent  conformal transformation for a Friedmann–Lema\^itre–Robertson-Walker (FLRW) metric, since it corresponds to a scale factor redefinition from the Einstein to the Jordan frame:
\be
g_{\rm E}=\Omega^2 \, g_{\J} \rightarrow a_{\E}=\Omega \, a_{\J}\,.
\ee
Under the above 
conformal transformation
the GR Lagrangian takes the form 
\bea
\Lg_{\rm GR}&=&\Omega^2 a^2_{\J}\Big[ h'^2- (\nabla h)^2\Big]\ ,\label{LGRJ}
\eea
from which we get the ``apparently'' modified GW propagation equation
\be
h''+2  {\h}_{\J}\Big(1+ \frac{\Omega'}{{\h}_{\J}\Omega}\Big) h'- \nabla^2 h=0  \,,
\ee
where $\h_{\J}=a_{\J}'/a_{\J}$.
Due to the conformal invariance of tensor modes, 
\be
{\h}_{\J}\Big(1+\frac{\Omega'}{{\h}_{\J}\Omega}\Big)=\h_{\E}~,\label{fric}
\ee
in agreement with the definitions of ${\h}_{\J}$ and ${\h}_{\E}$. 
Hence, the friction term as a function of space and conformal time, is the same one, just written in different frames related by conformal transformation.

The above arguments are completely general, and apply to any theory for which the GW propagation equation has luminal speed and a modified friction term, independently of the tensorial type and number of  additional physical degrees of freedom.

\textbf{GW propagation in MGTs---}
Consider 
MGTs with a scalar degree of freedom. The corresponding effective field theory (EFT) was formulated in the Jordan frame \cite{Gubitosi:2012hu,Gleyzes:2013ooa}, showing the relation to the Einstein frame EFT, known as effective field theory of inflation \cite{Cheung:2007st}.

Such an approach can be applied 
to the Horndeski theory, or other MGTs involving a scalar field. For the specific case of Horndeski theory, 
the transformation between Einstein and Jordan frames can be derived using the conformal invariance of tensor and curvature perturbations \cite{Romano:2023uwf}. Alternatively, without using perturbation theory, one can find the  relation between frames by studying the effects of conformal transformations on the  non-perturbed MGT field equation  \cite{Romano:2022jeh}, or by writing explicitly the Horndeski Lagrangian in the form of the quadratic Lagrangian for an effective field theory of DE in the Jordan frame \cite{Gubitosi:2012hu} %given in eq.(\ref{LDEJ}).
\bea
\Lg^{\rm eff}_{\rm DE}&=&\sqrt{g_{\J}}\Big[\Omega^2 R_{\J} + L^{(2)}_{\J}\Big]\,;\label{LDEJ}
\eea
$L^{(2)}_{\J}$ stands for all remaining terms.
Note that
the Einstein frame is defined as the frame in which the Lagrangian has the Hilbert form
\bea
\Lg^{\rm eff}_{\rm DE}&=&\sqrt{g_{\E}}\Big[R_{\E} + L^{(2)}_{\E}\Big]\,.
\eea
where $g_{\E}=\Omega^2 g_{\J}$. 

Following the EFT for DE, the Lagrangian for tensor perturbations in the Jordan frame is
\cite{Gleyzes:2013ooa}
\bea
\Lg^{\rm eff}_h&=&\frac{a_{\J}^2\Omega^2}{c_{\T}^2}\Big[h'^2-c_{\T}^2(\nabla h)^2\Big]\,, \label{LheffJ}
\eea
%while 
and in the Einstein frame \cite{Creminelli:2014wna}  is
\bea
\Lg^{\rm eff}_h&=&\frac{a_{\E}^2}{c_{\T}^2}\Big[h'^2-c_{\T}^2(\nabla h)^2\Big]\,, \label{LheffE}
\eea
where $a_{\E}=\Omega \, a_{\J}$, consistent with the general conformal transformation $g_{\E}=\Omega^2 g_{\J}$ defined for the full action.

A more general effective Lagrangian was derived in \cite{Romano:2022jeh}, 
valid for an arbitrary number 
of fields, and including the effects of higher order terms.

From Eq.(\ref{LheffE}),
the GW propagation equation in the Einstein frame reads
 \bea
h''+2 \Big(\frac{a_{\E}'}{a_{\E}}-\frac{c'_{\T}}{c_{\T}}\Big) h'-c_{\T}^2 \nabla^2 h=0\,.
\label{GWpr}
\eea
Clearly, for $c_{\T}'=0$ the friction term cannot be modified. In particular, if GWs propagate at the speed of light, Eq.(\ref{GWpr}) reduces to the  one of general relativity.

From the effective Lagrangian in Jordan frame, Eq.(\ref{LheffJ}), the GW propagation equation reads
\be
h''+2  {\h}_{\J}\Big(1-\frac{c_{\T}'}{{\h}_{\J} c_{\T}}+\frac{\Omega'}{{\h}_{\J}\Omega}\Big) h'-c_{\T}^2 \nabla^2 h=0  \,,\label{hctO}
\ee
where 
${\h}_{\J}=a_{\J}'/a_{\J}$.
For luminal gravitational waves, the above equation simplifies to
\be
h''+2  {\h}_{\J}\Big(1+\frac{\Omega'}{{\h}_{\J}\Omega}\Big) h'- \nabla^2 h=0  \,,\label{hctO1}
\ee
corresponding to the Lagrangian
\be
\Lg_{\rm GR} =a_{\J}^2\Omega^2\Big[ h'^2- (\nabla h)^2\Big]=a_{\E}^2\Big[ h'^2- (\nabla h)^2\Big]\ .\label{Lheff1}
\ee
Hence the EFT approach  confirms the result obtained 
previously,
namely that for GWs propagating at luminal speed, any phenomenological parametrisation of their propagation equation involving a modification of the friction term should have no physical relevance. Such modification corresponds to GR written in a different frame. 
In the following we will consider MGTs that can be described by the EFT formulated in \cite{Gleyzes:2013ooa}.

%%%%%%%%%%%%%%%%%%%%%%%%%%%%%%%%%%%%%%%%%%%%%%%%%%

\textbf{Luminosity distances---}
In Minkowski background, the GW amplitude $h$ is inversely proportional to the distance from the source $r$. 
To study the effects of cosmological expansion on $h$ it is convenient to write the Lagrangian for an expanding universe, 
Eq.(\ref{LheffE}),  as 
\be
\Lg^{\rm eff}_h=\alpha^2\Big[ h'^2-c_{\T}^2 (\nabla h)^2\Big]~,
\label{Lheffalpha}
\ee
where
\be
\alpha=\frac{a_{\E}}{c_{\T}}=\frac{\Omega a_{\J}}{c_{\T}}~,
\label{def}
\ee
leading to the propagation equation 
\be
h''+2 \frac{\alpha'}{\alpha} h'-c_{\T}^2 \nabla^2 h =0~; \label{halpha}
\ee
 $\alpha$ plays the role of an effective scale factor for the GW propagation.

Introducing a new parameter $\chi$ as $h=\chi/\alpha$,
Eq.(\ref{halpha})
in Fourier space (denoted by a subscript $k$) reads\footnote{This is consistent with Eq.(15) in \cite{Belgacem:2018lbp} for $c_{\T}=1$.
The definition is conformally invariant, since both $r$ and $\alpha$ are conformally invariant. This is expected, since $h$ is conformally invariant.}
\be
\chi_k''+\Big(c_{\T} k^2 -\frac{\alpha''}{\alpha}\Big)\chi_k=0 \,.
\ee
On sub-horizon scales $\alpha''/\alpha$ can be neglected,
implying
\be
h_k \propto \frac{1}{\alpha} \,.
\ee
Hence, on sub-horizon scales GW amplitude in an expanding Universe evolves as
\bea
h_k
&\propto& \frac{1}{\alpha \, r} \propto 
 \frac{1}{d^{\rm GW}_{\rm L}}\,, 
\eea
where
\be
d^{\rm GW}_{\L}(z)=r \frac{\alpha(0)}{\alpha(z)} \,,\label{dgwalpha}
\ee
stands for the gravitational luminosity distance\cite{Maggiore:2007ulw}, a quantity inferred from GW observations\footnote{
The definition Eq.(\ref{dgwalpha}) is not equivalent to the one given in terms of the flux of gravitational radiation \cite{Lobato:2022puv}.}.  For a flat FLRW background  
the comoving distance $r(z)$ is 
 \be
 r(z)=\int^z_0\frac{{\rm d}z'}{H(z')} \,. \label{rfrw}
 \ee
 Note that Eq.(\ref{dgwalpha}) is valid in both the Einstein and the Jordan frames.

Let us denote with $\eta_{\rm s}$ and $\eta_{\rm o}$ the conformal time at the source and observer, respectively, and also denote with $d^{\rm EM}|_{\E}, d^{\rm EM}|_{\J}$ and $d^{\rm GW}|_{\E}, d^{\rm GW}|_{\J}$ the  EMW and GW luminosity distance of theories with matter coupled respectively to the Einstein and Jordan frame metrics. 
At leading order in perturbations, the distinction is not physically relevant for GWs, since independently of the matter-gravity coupling, 
\be
d_{\L}^{\rm GW}|_{\E}=d_{\L}^{\rm GW}|_{\J}=r\, \frac{\alpha(\eta_{\rm o})}{\alpha(\eta_{\rm s})}=\frac{c_{\rm T}(\eta_{\rm s})}{c_{\rm T}(\eta_{\rm o})}d^{\rm GR}_{\L}\,,
\ee
where $d^{\rm GR}_{\L}$ corresponds to Eq.(\ref{dgwalpha}) with $c_T=c$, as shown in Table I.
However, for EMWs it is important, since it affects the definition of red-shift, and $d_{\L}^{\rm EM}|_{\E}\neq d_{\L}^{\rm EM}|_{\J}$. 

\begin{table*}[t]
\centering
\begin{tabular}{ | m{2.5cm} | m{3.5cm}| m{4.5cm} |  m{5.5cm} | }
\hline
  &GR & MGT - Einstein frame & MGT - Jordan frame    \\ \hline
 Scale factor & $a_{\E}$ & $a_{\E}$ & $a_{\J}={(M_* c_{\T})}^{-1} a_{\E}$ \\ \hline
 Coupling	& $g_{\E}$ & $g_{\E}$ & $g_{\J}$ \\ \hline
 $(1+z)$  & $\frac{a_{\E}(0)}{a_{\E}(z)}$ & $\frac{a_{\E}(0)}{a_{\E}(z)}$  &  $\frac{a_{\J}(0)}{a_{\J}(z)}=\frac{M_*(z) c_{\T}(z) a_{\E}(0)}{M_*(0) c_{\T}(0) a_{\E}(z)}$\\  \hline
 $d^{\rm GW}_L$ & $d^{\rm GW}_{\rm GR}=r\frac{a_{\E}(0)}{a_{\E}(z)}$ & $d^{\rm GW}_{\rm MGT}=\frac{c_{\T}(z)}{c_{\T}(0)}d^{\rm GW}_{\rm GR}$   & $d^{\rm GW}_{\rm MGT}=\frac{c_{\T}(z)}{c_{\T}(0)}d^{\rm GW}_{\rm GR}$ \\ \hline
$d^{\rm EM}_L$ & $d^{\rm EM}_{\rm GR}=r\frac{a_{\E}(0)}{a_{\E}(z)}$ & $d^{\rm EM}_{\rm MGT}=d^{\rm EM}_{\rm GR}$   & $d^{\rm EM}_{\rm MGT}=r\frac{a_{\J}(0)}{a_{\J}(z)}=\frac{M_*(z) c_{\T}(z)}{M_*(0) c_{\T}(0) }d^{\rm EM}_{\rm GR}$ \\ \hline
$\frac{d^{\rm GW}_{\L}}{d^{\rm EM}_{\L}}$ & 1& $\frac{c_{\T}(z)}{c_{\T}(0)}$ & $\frac{M_*(0)}{M_*(z)}$ \\ \hline
% $d_L{\}$  &  $d_{\L}^{\rm GW}|_{\E}=\frac{c_T(z)}{c_{\T}(0)}  d_L^{\rm EM}(z)$ & d_L^{\rm GW}|_{\J}=\frac{M_*(0)}{M_*(z)}d_{\L}^{\rm EM}   \\ \hline
\end{tabular}
\caption{Redshift, gravitational  and electromagnetic luminosity distance relations for GR and MGTs with Einstein or Jordan frame matter-gravity couplings. 
}
\label{tabI}
\end{table*}

\textbf{Conformal transformations and redshift-scale factor relation---}
The Lagrangian of electromagnetism is conformally invariant  \cite{Cote:2019kbg}, but the relation between redshift and scale factor can change, depending on which metric $A^{\mu}$ is coupled to.
This can be understood using the geometric optical approximation \cite{Ellis:1971pg}. We give below a proof,  using the invariance of the norm of the photon four-momentum.

Consider  a conformal transformation, and its effect on the scale factor of the FLRW metric
\bea
g_{\E}=\Omega^2 g_{\J} &,& a_{\E}=\Omega \, a_{\J}\nonumber \,.
\eea
The norm of the four-momentum of a photon must be zero in any frame, but if photons are minimally coupled with the metric $g_{\J}$, i.e. indices are contracting with $g_{\J}$, we have
\bea
P_{\mu}P^{\mu}&=&P^{\mu}P^{\nu}g_{\mu\nu}^{\J}\nonumber\\&=&E^2-\delta_{ij}p^i p^ja_{\J}^2\nonumber\\
&=&E^2-\delta_{ij}p^i p^j \Omega^{-2} a_{\E}^2 \nonumber\\
&=&0 \nonumber\,;
\eea
$p^j$ are the components of the comoving momentum. Hence,
\bea
E&=& p ~a_{\J}= p~ a_{\E} \Omega^{-1}~,\nonumber
\eea
with $p^2=\delta_{ij}p^i p^j$ the norm of the comoving momentum.
We can then compute the redshift as
\bea
(1+z)&=&\frac{E(z)}{E(0)}=\left[\frac{a(0)}{a(z)}\right]_{\J}=\frac{\Omega(z)}{\Omega(0)}\left[\frac{a(0)}{a(z)}\right]_{\E}\nonumber \,,
\eea
showing that the relation between redshift and scale factor depends on which metric is minimally coupled to the matter fields, namely with which metric tensor indices are raised and lowered.

\textbf{Luminal modified gravity theories---}
As shown previously,
for $c_{\T}=1$, Eq.(\ref{hctO}) implies that  
the only way to get an apparent modification of the friction term is to write the GR action in the Jordan frame.
From the definition of gravitational luminosity distance, Eq.(\ref{dgwalpha}), 
\be
d_ {\L}^{\rm GW}|_{\E}
=
r \frac{\alpha(0)}{\alpha(z)}
=
r \frac{a_{\E}(0)}{a_{\E}(z)}
=r(1+z)~.
\ee
Using the definition
\be
d_{\L}^{\rm EM}=r(1+z)~,
\label{dlem}
\ee
valid in either the Einstein or the Jordan frame, we get
\be
d_ {\L}^{\rm GW}|_{\E}
=d_{\L}^{\rm EM}|_{\E}~.
\ee
So far we have assumed
\be
(1+z)=\frac{a_{\E}(0)}{a_{\E}(z)} \,,
\label{1+z}\ee
namely that matter fields are minimally coupled to the metric tensor in the Einstein frame $g_{\E}$. 

If matter fields are minimally coupled to the metric tensor in the Jordan frame $g_{\J}$, we get the following redshift scale factor relation \cite{Ellis:1971pg}:
\be
%a_E&=&\Omega \,a_J \\
(1+z)=\frac{a_{\J}(0)}{a_{\J}(z)}=\frac{a_{\E}(0)}{a_{\E}(z)}\frac{\Omega(z)}{\Omega(0)}~,  \label{zJM}
\ee 
and consequently
\be  
d_{\L}^{\rm EM}|_{\J}=r (1+z)=r \frac{a_{\E}(0)}{a_{\E}(z)}\frac{\Omega(z)}{\Omega(0)}=\frac{\Omega(z)}{\Omega(0)} d_{\L}^{\rm GW}|_{\J} \,, \label{dEMJ}
\ee
in agreement with \cite{Belgacem:2018lbp,Dalang:2019rke}.

Within the context of modified gravity theories, it is customary to interpret $\Omega$ as an effective Planck constant\footnote{Note that it is the quantity $\Omega=M_* c_{\T}$ that plays the role of an effective Planck mass, since it is the coefficient of the Ricci scalar in Eq.(\ref{LDEJ}). In the case of $c_{\T}=c$, then $\Omega=M_*$.}
 $M_*$. Hence,
\be  
d_{\L}^{\rm EM}|_{\J}=\frac{M_*(z)}{M_*(0)}d_{\L}^{\rm GW}|_{\J} \,. \label{dfine}
\ee
The relation between $d_{\L}^{\rm EM}|_{\rm E}$ in the Einstein frame and $d_{\L}^{\rm EM}|_{\J}$ in the Jordan frame is
\be
d^{\rm EM}_{\L}|_{\J}=r\frac{a_{\J}(0)}{a_{\J}(z)}=r \frac{a_{\E}(0)}{a_{\E}(z)}\frac{M_*(z)}{M_*(0)} =\frac{M_*(z)}{M_*(0)}d_{\L}^{\rm EM}|_{\E}\, .
\ee
% This is expected since while the electromagnetic Lagrangian density is conformally invariant \cite{Cote:2019kbg}, the vector potential $A^{\mu}$ is not, leading to $d_{\L}^{\rm EM}|_{\J} \neq d_{\L}^{\rm EM}|_{\E}$.

Regarding the GW luminosity distance, from Eq.(\ref{dgwalpha}) with the definition Eq.(\ref{def}), we get
\be
d_{\L}^{\rm GW}|_{\J}=d_{\L}^{\rm GW}|_{\E}~. 
\ee
Hence, $d_{\L}^{\rm GW}$ is not affected by $M_*$, as expected from the conformal invariance of $h$.

The relation between the GW and 
EMW luminosity distances, Eq.(\ref{dfine}), holds for matter fields minimally coupled to the metric tensor in the Jordan frame.
As a result, a GW-EMW luminosity distance relation of the kind obtained in \cite{Belgacem:2018lbp}, should be interpreted as the effect of the propagation of EMWs within a theory of electromagnetism where the Jordan frame  $g_{\J}$ (and not $g_{\E}$) is the metric coupled to the vector potential.
In other words, if $c_{\T}=1$, gravitons redshift in the same way as in GR, because they do not feel the effective Planck mass, but if photons are coupled to $g_{\J}$ they do feel it, leading to a difference between the gravitational and electromagnetic luminosity distances.

\textbf{Non-luminal modified gravity theories---}
The effective Jordan frame Lagrangian reads
\cite{Lagos:2019kds}
\bea
\Lg^{\rm eff}_h&=&M_*^2 a_{\J}^2\Big[ h'^2-c_{\T}^2 (\nabla h)^2\Big] \,.
\eea
which compared to the Lagrangians Eq.(\ref{LheffE}) and (\ref{LheffJ})  implies the  frame transformation
\bea
a_{\E}&=&M_* c_{\T} \,a_{\J}=\Omega \,a_{\J} \,, \label{aej}
\eea
from which we get
\bea
\alpha&=&\frac{a_{\E}}{c_{\T}}=M_* {a_{\J}} \,.
\eea
From the relation between frames we can derive the model-independent relations for the non-luminal modified gravity theories, summarised in Table~I.

For Einstein frame photon-graviton coupling, using Eqs.(\ref{dlem}), (\ref{1+z}), 
\bea
d_{\L}^{\rm EM}|_{\E}&=&r(1+z)=r\frac{a_{\E}(0)}{a_{\E}(z)} \,, \\
d_{\L}^{\rm GW}|_{\E}&=&r \frac{\alpha(0)}{\alpha(z)}= r \frac{a_{\E}(0)}{a_{\E}(z)} \frac{c_{\T}(z)}{c_{\T}(0)}=\frac{c_{\T}(z)}{c_{\T}(0)}d_L^{\rm EM}|_{\E}\label{dgw} \,.
\eea
Setting $c_{\T}(z)=1$, we get $d_{\L}^{\rm GW}|_{\E}=d_{\L}^{\rm EM}|_{\E}$, in agreement with our previous result.

For Jordan frame photon-graviton coupling, we obtain
\bea
d_{\L}^{\rm EM}|_{\rm J}&=&r(1+z)=r\frac{a_{\J}(0)}{a_{\J}(z)}=r \frac{a_{\E}(0)}{a_{\E}(z)}
\frac{M_{*}(z)}{M_{*}(0)}
\frac{c_{\T}(z)}{c_{\T}(0)}\nonumber
\\&=&\frac{M_{*}(z)}{M_{*}(0)}
\frac{c_{\T}(z)}{c_{\T}(0)}
d_{\L}^{\rm EM}|_{\rm E}~,
\eea
and
\bea
d_{\L}^{\rm GW}|_{\rm J}&=&
r\frac{\alpha(0)}{\alpha(z)}=r\frac{M_*(0)}{M_*(z)}\frac{ a_{\J}(0)}{a_{\J}(z)}\nonumber\\&=&
\frac{M_*(0)}{M_*(z)}
d_{\L}^{\rm EM}|_{\J}~,
\eea
hence
Eq.(\ref{dfine}) is also valid for non-luminal modified gravity theories.

\textbf{Observational implications---}
To understand the observational implications of the results presented above, it is important to distinguish between the frame one uses to perform calculations and the matter-gravity coupling frame. 
Observational quantities do not depend on the choice of the frame one uses to perform their calculations. On the contrary, the choice of the frame for the coupling between  matter and gravity may lead to observational implications, since this choice corresponds to the selection of a different Lagrangian.

The results summarized in Table~I  do not depend on the frame one uses to perform  calculations, however the difference between the third and fourth columns corresponds to two different families of theories, and is indeed observable.

For luminal MGTs the gravitational luminosity distance is the same as in GR for any matter-gravity coupling, while the electromagnetic luminosity distance is modified only for Jordan frame coupling.
For non-luminal MGTs the gravitational luminosity distance is modified with respect to GR, and is the same for any matter-gravity coupling, while the electromagnetic luminosity distance is modified only for Jordan frame coupling.
Independent GW and EMW observations are necessary to constrain the effects  on the GW and EMW luminosity distances.

%%%%%%%%%%%%%%
\textbf{Conclusions---}
%%%%%%%%%
 A modification of the friction term of the GW propagation equation, in the context of a modified gravity theory, leads to a (real) physical effect only if gravitational waves do not propagate at luminal speed. Otherwise, the propagation of  GWs and the gravitational luminosity distance are the same as within GR. 
 
 Studying modified gravity theories in the Jordan frame, as it is customary, one observes an apparent modification in the GW propagation equation, even for GWs propagating at luminal speed. This however has no physical effect due to conformal invariance.

 The GW luminosity distance differs from the EMW luminosity distance even for modified gravity theories with GWs propagating at luminal speed, if photons are coupled to the Jordan frame metric, instead of the Einstein frame metric.
 In this case, the EMW luminosity distance is different from the corresponding expression in GR, despite that the GW propagation remains unaffected.

Hence, the obtained GW-EMW luminosity distance relation modification obtained when gravitational waves propagate at the speed of light, may be regarded in the Einstein frame as the manifestation of the modification of electromagnetism, rather than a modification of gravity in the Jordan frame.

The model independent GW-EMW luminosity distance relation obtained for Einstein frame coupling, can be useed to analyse GWs and EMWs emitted by  bright sirens, since it allows to reconstruct $c_{\T}(z)$ from the GW and EMW luminosity distances.

\begin{acknowledgments}

We thank Tessa Baker,  Timothy Clifton, Charles Dalang, Martin Kunz,  Lucas Lombriser, Claudia de Rham, Isabela Santiago de Matos,  Michele Maggiore, Juan Maldacena, and Sergio Vallejo and Jorge Noreña for discussions. 
\end{acknowledgments}

\bibliographystyle{apsrev4-1}
\bibliography{Bibliography}

\end{document}